\documentclass[ reprint,
 aps,
 pra,
 showkeys,
 nofootinbib
]{revtex4-1}
\usepackage{amsfonts}
\usepackage{amssymb}
\usepackage{amsmath}
\usepackage{graphicx}
\usepackage{subfigure}
\usepackage{dcolumn}
\usepackage{bm}
\usepackage{setspace}
\usepackage[hang]{footmisc}
\usepackage[sort&compress]{natbib}
\usepackage{multirow}
\usepackage[sort&compress]{cleveref}
\usepackage{nomencl}
\usepackage[english]{babel}
\usepackage{xcolor, soul}
\usepackage{marginnote}
\usepackage{verbatim}
\sethlcolor{yellow}
\usepackage{ragged2e}

\crefname{equation}{\unskip}{\unskip}
\crefname{figure}{\unskip}{\unskip}
\crefname{section}{\unskip}{\unskip}
\crefname{subsection}{\unskip}{\unskip}
\begin{document}
\let\ref\cref

\title{Cylinder--flat contact mechanics during sliding}

\author{J. Wang}
\affiliation{PGI-1, FZ J\"ulich, Germany, EU}
\affiliation{College of Science, Zhongyuan University of Technology, Zhengzhou 450007, China}
\author{ A. Tiwari}
\affiliation{PGI-1, FZ J\"ulich, Germany, EU}
\affiliation{www.MultiscaleConsulting.com}
\author{ I. Sivebaek}
\affiliation{PGI-1, FZ J\"ulich, Germany, EU}
\affiliation{Department of Mechanical Engineering, Technical University of Denmark,
        Produktionstorvet, Building 427, Kongens Lyngby 2800, Denmark}
\affiliation{Novo Nordisk Device R \& D, DK-400 Hiller$\phi$d, Denmark}
\author{ B.N.J. Persson}
\affiliation{PGI-1, FZ J\"ulich, Germany, EU}
\affiliation{www.MultiscaleConsulting.com}

\begin{abstract}
Using molecular dynamics (MD) we study the dependency of the contact mechanics on the sliding speed when an elastic
block (cylinder) with a ${\rm cos} (q_0 x)$ surface height profile is sliding in adhesive contact on a rigid flat substrate.
The atoms on the block interact with the substrate atoms by Lennard-Jones (LJ) potentials, and we consider both commensurate and
(nearly) incommensurate contacts. For the incommensurate system the friction force
fluctuates between positive and negative values, with an amplitude proportional to the sliding speed,
but with the average close to zero. For the commensurate system the (time-averaged) friction force is much
larger and nearly velocity independent. For both type of systems the width of the contact region is velocity independent even when,
for the commensurate case, the frictional shear stress increases from zero (before sliding) to
$\approx 0.1 \ {\rm MPa}$ during sliding. This frictional shear stress,
and the elastic modulus used, are typical for Polydimethylsiloxan (PDMS)
rubber sliding on a glass surface, and we conclude
that the reduction in the contact area observed in some experiments when increasing
the tangential force must be due to effects not included in our model study, such as viscoelasticity or elastic nonlinearity.
\end{abstract}

\maketitle
\makenomenclature

%%%%%%%%%%%%%% main text %%%%%%%%%%%%%%%%
%\begin{multicols}{2}

\begin{figure}
\centering
\includegraphics[width=0.45\textwidth]{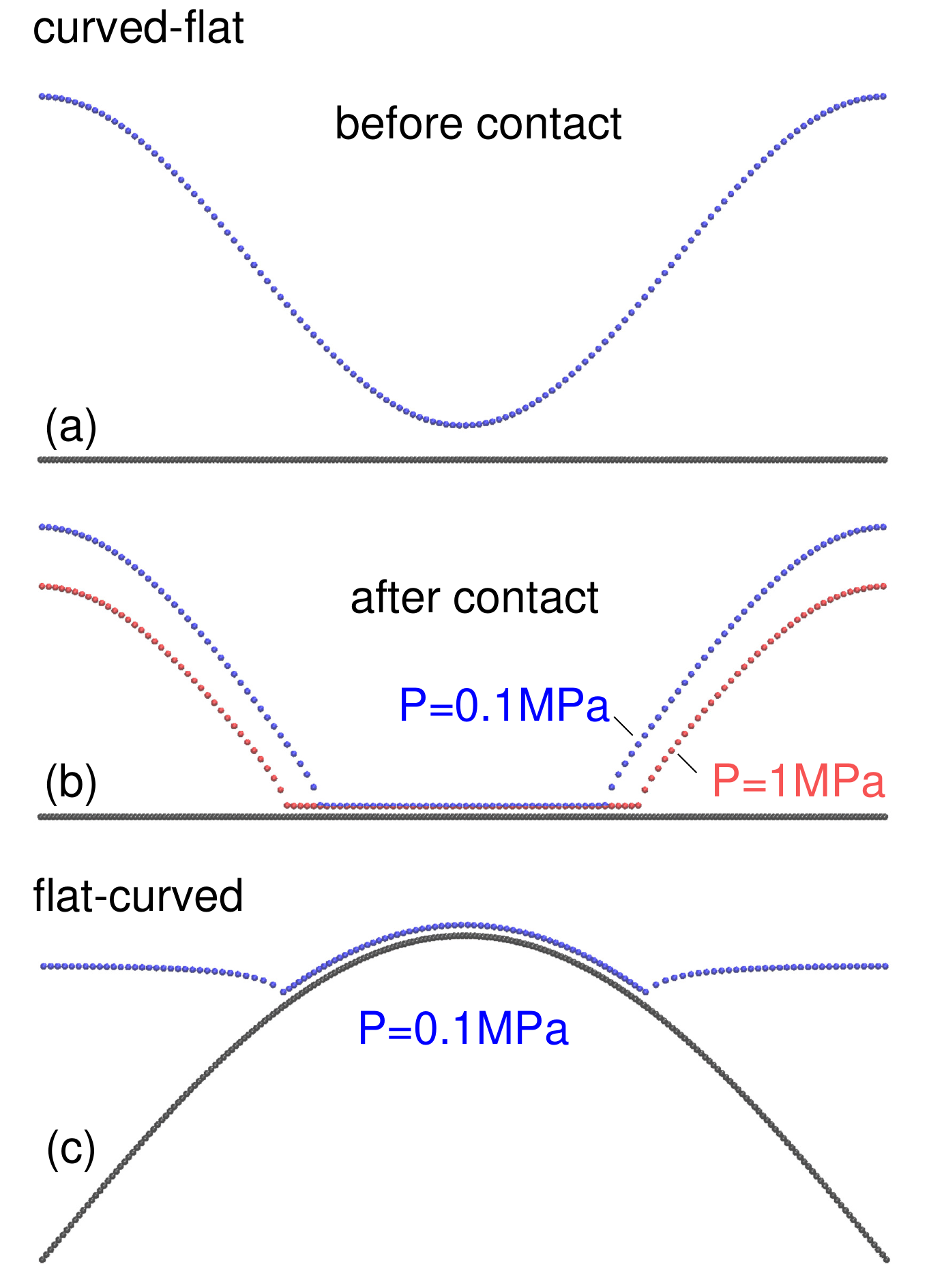}
\caption{\label{NewCombined.pdf}
The contact area between an elastic slab (block) and a rigid substrate at the temperature $T=0 \ {\rm K}$.
In (a) and (b) the block is corrugated with the height
coordinate $z=h_0 {\rm cos} (q_0 x)$ with $q_0=2 \pi/L_x$.
In (c) the substrate has the same corrugation amplitude but with double wavelength
(i.e. $q_0=\pi /L_x$), while the block has a flat surface.
We denote the two different systems as curved-flat and flat-curved.
For the curved-flat system we show the contact (a) before, and (b) after squeezing the block against the
substrate with the nominal contact pressure $p= 0.1 \ {\rm MPa}$ (blue), and $p=1 \ {\rm MPa}$ (red).
For the flat-curved system we show the contact for the nominal contact pressure $p= 0.1 \ {\rm MPa}$.
The Young's modulus for the elastic block $E=10 \ {\rm MPa}$,
and the LJ block-substrate atom interaction parameters are given in the text.}
\end{figure}

\centering{\bf INTRODUCTION}
\justify
The contact between a spherical (or cylindrical) body and a flat surface is
perhaps the simplest possible contact mechanics problem,
and often used in model studies of adhesion and friction\cite{AviDoroBo,AviDoroBo1,AviDoroBo2}.
For stationary contact with $F_x=0$, where $F_x$ is the applied tangential force,
the adhesive interaction is well described by the Johnson-Kendall-Roberts (JKR) theory\cite{JKR,Jons} which has been tested in great detail.
However, when the tangential force $F_x$ is non-zero the problem becomes much more complex
and not fully understood\cite{Savkoor,Jon,Menga,Kim,a3,a6}.

Here we consider the contact between
an elastic block with cylinder shape with the height profile $z=h_0 {\rm cos} (q_0 x)$, and a rigid solid with a flat surface.
We will refer to this system as curved-flat. In Ref. \cite{a3} and \cite{a6}
we studied the opposite situation of an elastic block with a flat surface
in contact with a rigid solid with the height profile $z=h_0 {\rm cos} (q_0 x)$. We will refer to this system as flat-curved.
When $F_x = 0$ the curved-flat and flat-curved systems are both described by the JKR theory.
However, as will be shown here, during sliding the two systems exhibit very different properties.

\vskip 0.3cm
\centering{\bf MODEL}
\justify
A curved elastic block can be obtained by ``gluing'' an
elastic slab to a rigid upper surface profile. Here we use a slab
of thickness $L_z \approx 86  \ {\rm \AA}$ attached to a rigid surface with the
height profile $z=h_0 {\rm cos} (q_0 x)$, where $h_0 = 100  \ {\rm \AA}$ and $q_0 =  2 \pi /L_x$.
We use periodic boundary conditions in the $xy$-plane with $L_x= 254 \ {\rm \AA}$ and $L_y=14 \ {\rm \AA}$.
The number of atoms in the $x$-direction is $N_x =128$ for the block, and for the substrate we consider two cases where
$N_x =128$ (commensurate interface) and $N_x = 206$. In the latter case the ratio between
the lattice constant of the block and the substrate is $a_{\rm b}/a_{\rm s} =206/128 \approx 1.609$,
which is close to the golden mean $(1+\surd 5)/2\approx 1.618$, i.e. the interface is nearly incommensurate.
The elastic block is treated using the smart-block description (with 13 layers with the same spacing as for the first layer
plus 4 layers on top of it, where at every step we double the lattice spacing in the $x$ and $z$-directions)
discussed in Ref. \cite{a3}, where the bending and elongation
spring constants are chosen to give the Young's modulus and the Poisson ratio
$E=10 \ {\rm MPa}$ and $\nu = 0.5$, respectively.
The interaction potential between the block and wall atoms at the interface is of the Lennard-Jones (LJ) type:
$$V(r) = 4 V_0 \left [\left ({r_0\over r}\right )^{12}- \left ({r_0\over r}\right )^{6}\right ],$$
where $V_0 = 0.04 \ {\rm meV}$ and $r_0=3.28 \ {\rm \AA}$. With this interaction potential we calculate the (adiabatic) work of
adhesion $w\approx 0.0023 \ {\rm J/m^2}$ for the commensurate interface, and $w\approx 0.0027 \ {\rm J/m^2}$ for the incommensurate interface.
We note that in the present system, when the adhesion is removed ($w\rightarrow 0$) the contact width decreases
from $85.3 \ {\rm \AA} $ to $25.8 \ {\rm \AA}$ when the nominal pressure
$p= 0.1 \ {\rm MPa}$, and from $101.1 \ {\rm \AA}$ to $61.5 \ {\rm \AA}$ when $p= 1 \ {\rm MPa}$. Thus, in spite of the small
work of adhesion, the adhesion interaction is very important, which is due to the small size of the system (in the JKR theory
the width of the contact region depends on the
(dimensionless) parameter $w/(pR)$, where $R$ is a length characterizing the size of the system; $w/(pR)$ is
of order unity in the present case).

Fig \ref{NewCombined.pdf}
shows for the incommensurate interface the system (a) before contact, and (b)
after squeezing the block against the substrate with the nominal contact
pressure $p= F_z /(L_x L_y) = 0.1 \ {\rm MPa}$ (blue), and $p=1 \ {\rm MPa}$ (red).
We also show the contact for the flat-curved case studied in Ref. \cite{a3} and \cite{a6}.

\begin{figure}
\centering
\includegraphics[width=0.45\textwidth]{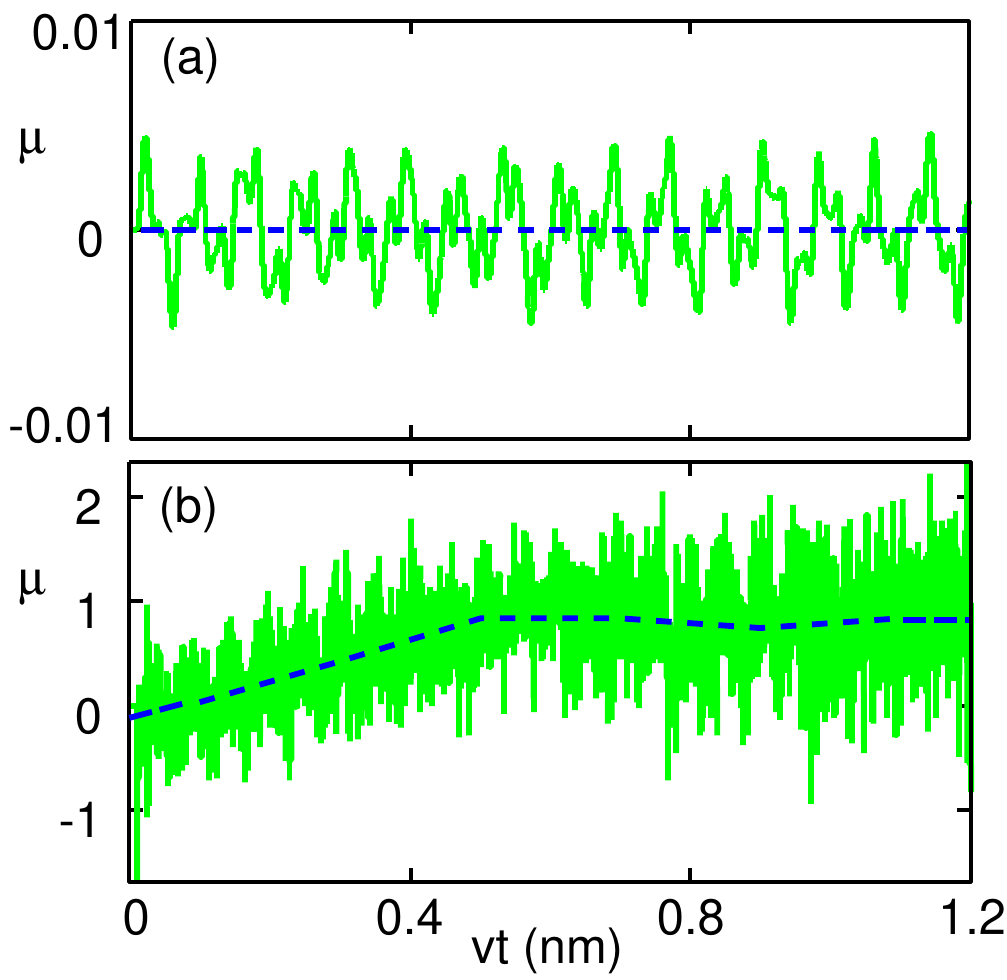}
\caption{\label{Comp.fri.comm-incomm-0.1Mpa-0.1ms-b.pdf}
The friction coefficient $\mu = F_x/F_z$ as a function of the distance $vt$ moved by the top of
the elastic block (curved-flat system).
For the incommensurate (a) and the commensurate (b) interface.
The nominal contact pressure $p=0.1 \ {\rm MPa}$ and the speed $v=0.1 \ {\rm m/s}$.}
\end{figure}

\begin{figure}
\centering
\includegraphics[width=0.45\textwidth]{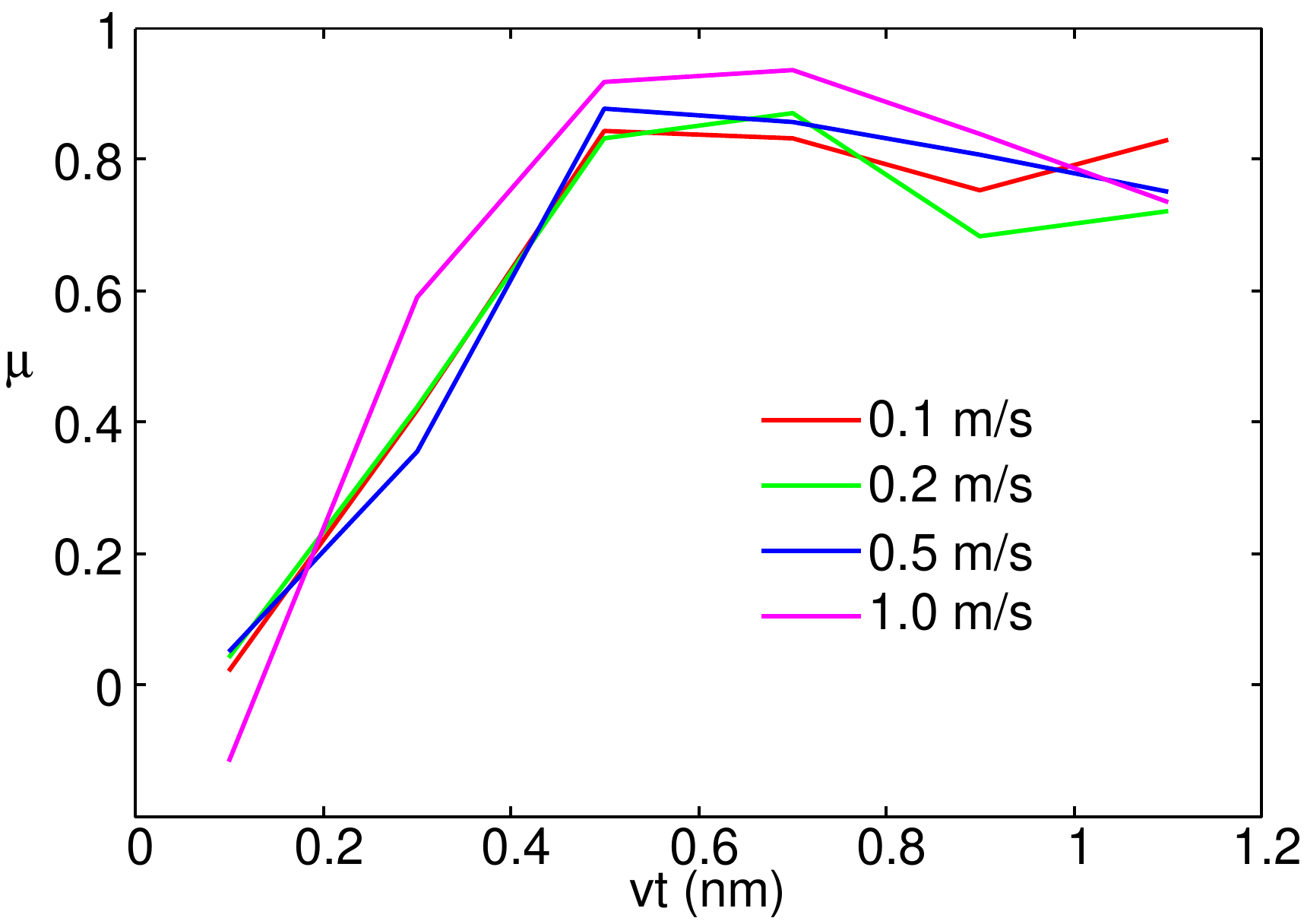}
\caption{\label{Fig.6a-commensurate-com-0.1-1.pdf}
The friction coefficient $\mu = F_x/F_z$  as a function of the
sliding distance for the commensurate system.
The sliding speed $v = 0.1, 0.2, 0.5$ and $1 \ {\rm m/s}$, and the contact pressure $p= 0.1 \ {\rm MPa}$.}
\end{figure}

\begin{figure}
\centering
\includegraphics[width=0.45\textwidth]{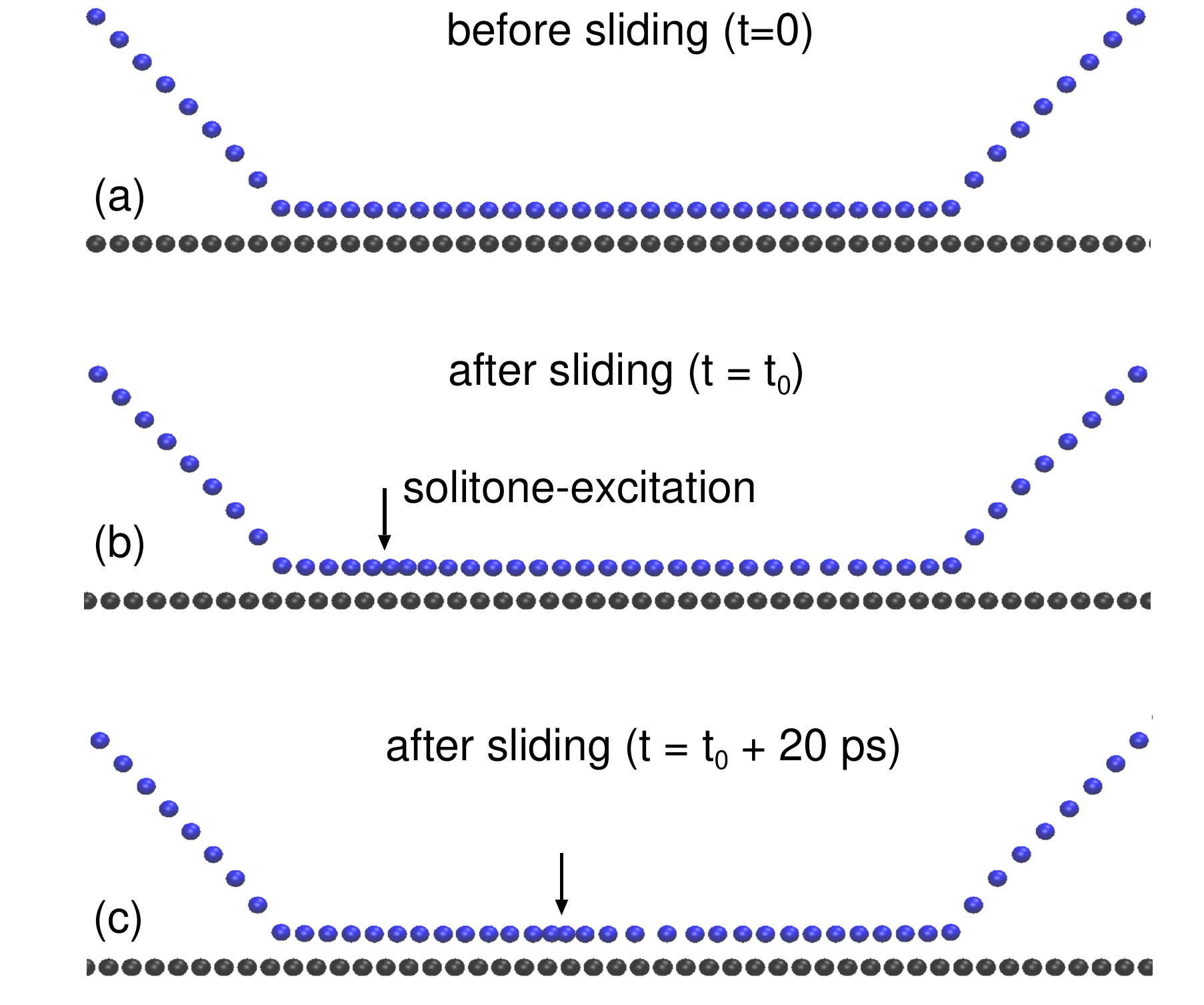}
\caption{\label{commensurate-sliding-detail.pdf}
For the curved-flat system with commensurate interface the sliding motion consist of
``long'' time periods of no sliding followed by rapid slip events where the block moves forward
by one substrate lattice spacing. The rapid motion consist of a compression domain wall (solitone-like
excitation) which propagate with a velocity ($\approx 43 \ {\rm m/s}$) close to the Rayleigh sound velocity
($\approx 0.95 c_{\rm T} \approx 53 \ {\rm m/s}$). During this rapid motion elastic waves (phonons) are radiated into the
block which is the origin of the observed friction force.}
\end{figure}

\begin{figure}
\centering
\includegraphics[width=0.45\textwidth]{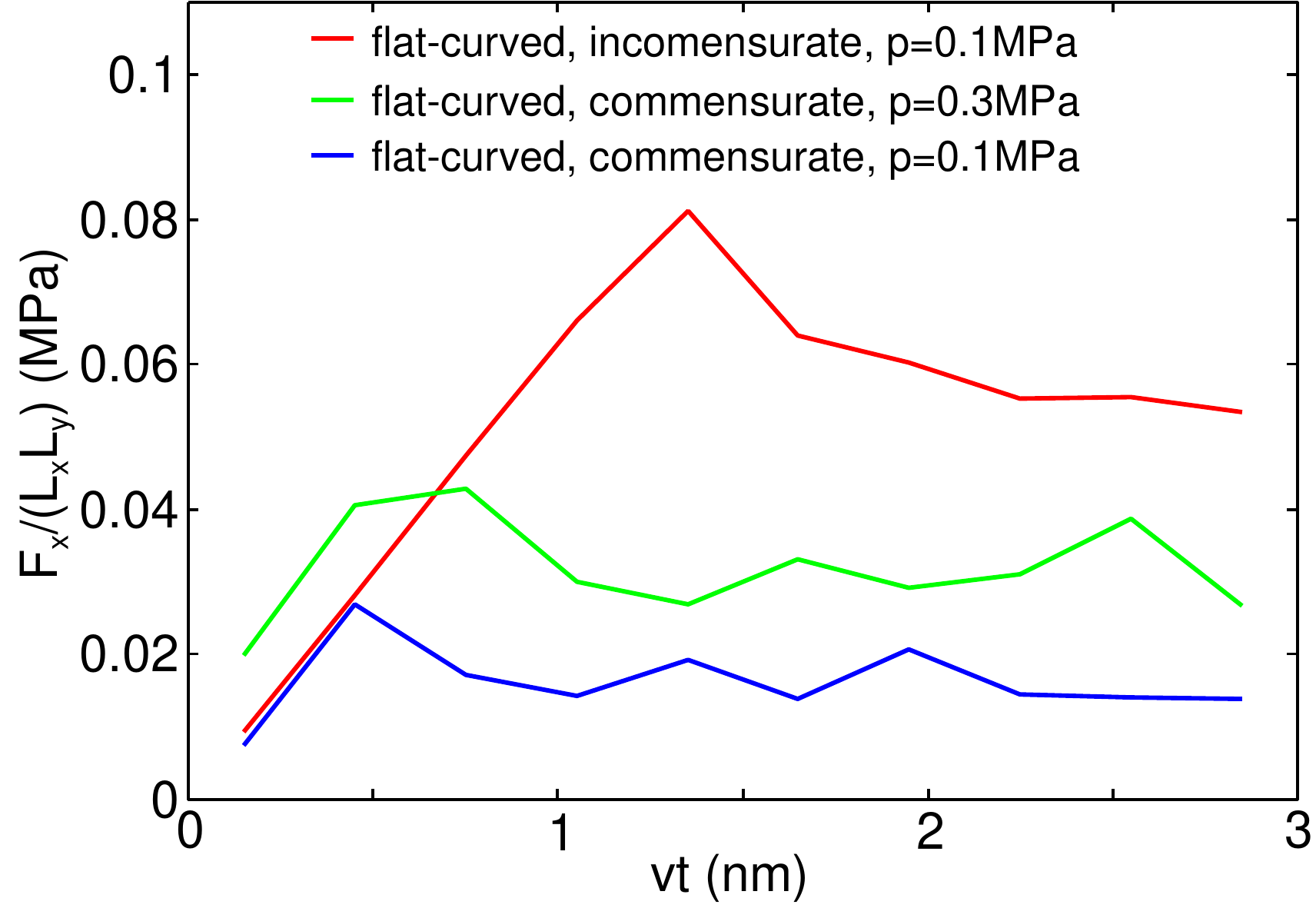}
\caption{\label{Fig.8-shear-stress-comm1x1-bendsub-0.1MPa-0.1ms.pdf}
The nominal shear stress as a function of the sliding distance
for the flat-curved system with different communicability's for the sliding speed $v=0.1 \ {\rm m/s}$}
\end{figure}

\begin{figure}
\centering
\includegraphics[width=0.45\textwidth]{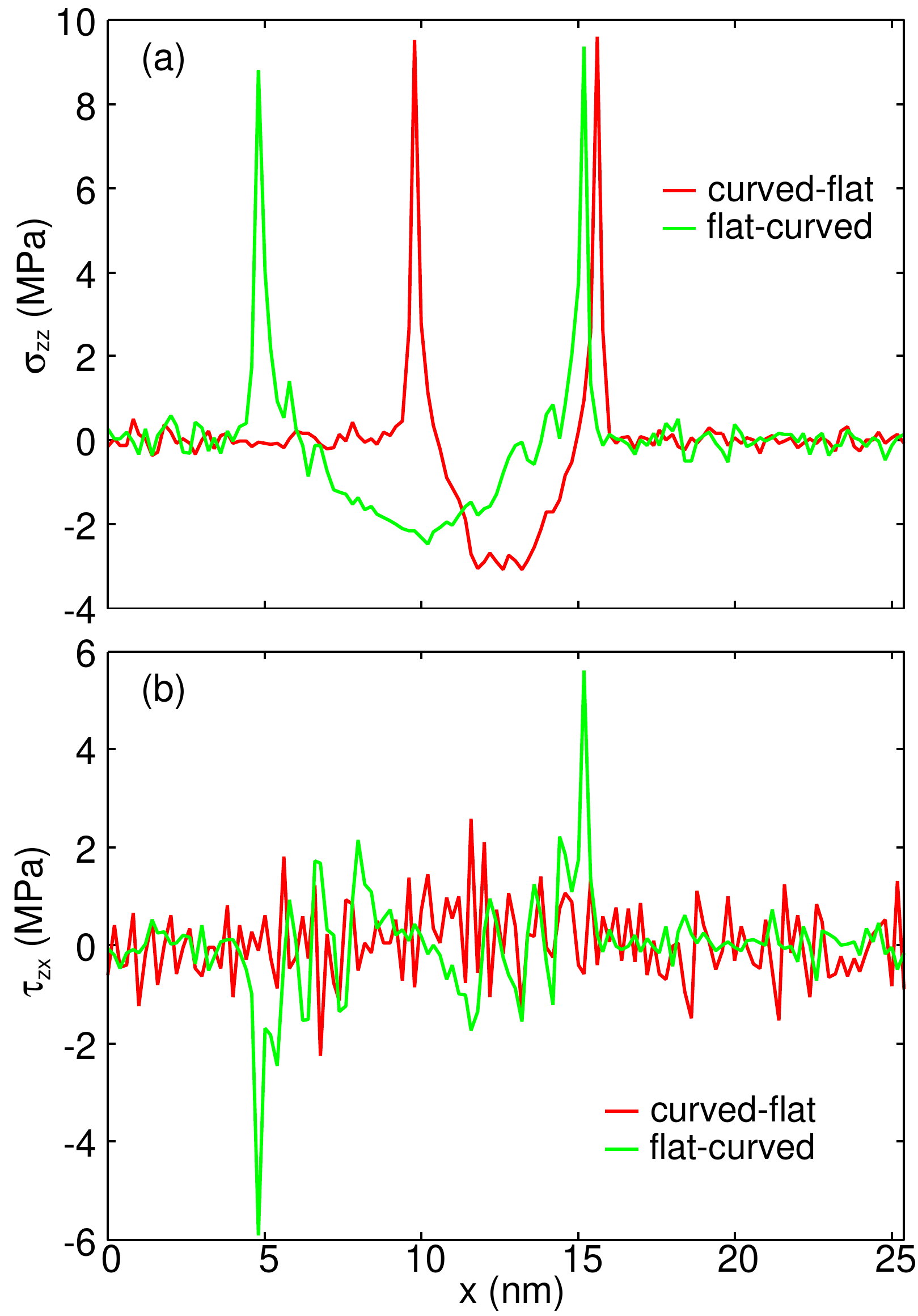}
\caption{\label{Fig.9-Shear.stress-comp-comm-slid-0.1MPa-0.1ms-z.x.pdf}
The (a) normal stress $\sigma_{zz}$ and (b) the shear stress $\tau_{zx}$
acting on the block as a function of the spatial coordinate $x$. For the curved-flat system
(red curves), and for the flat-curved system (green curves), in both cases with commensurate interface.
The sliding speed $v=0.1 \ {\rm m/s}$, and the nominal contact pressure $p=0.1 \ {\rm MPa}$.}
\end{figure}

\vskip 0.3cm
\centering{\bf NUMERICAL RESULTS}
\justify
Fig. \ref{Comp.fri.comm-incomm-0.1Mpa-0.1ms-b.pdf} shows the friction coefficient
$\mu = F_x/F_z$ (green lines) and its (local) average (blue) as a function of the distance
$s=vt$ moved by the upper surface of the block
for (a) the (nearly) incommensurate and (b) the commensurate system. The sliding speed $v=0.1 \ {\rm m/s}$ and
the nominal contact pressure $p= 0.1 \ {\rm MPa}$. For the incommensurate system the average friction coefficient
nearly vanish ($\mu < 10^{-4}$) while for the commensurate system it is of order unity ($\mu \approx 0.9$). The oscillations in
$F_x/F_z$ for the incommensurate system is due to the abrupt start of sliding where the upper surface of the block abruptly
start to move with the speed $v=0.1 \ {\rm m/s}$ at time $t=0$(see movie \cite{MovieA}). This result in an elastic wave propagating
(with the transverse sound velocity $c_{\rm T}= (G/\rho)^{1/2} \approx 56 \ {\rm m/s}$) towards
the interface so that only after the time $t=d/c_{\rm T}$ the atoms at the interface will start to move.
The periodicity of the fluctuation in the tangential force $F_x$ in Fig.  \ref{Comp.fri.comm-incomm-0.1Mpa-0.1ms-b.pdf}(a) is
given by the time it takes for an elastic wave to propagate back and forth between the two surfaces, i.e., the distance $2d$
giving the time $\Delta t=2 d/c_{\rm T}$ or sliding distance $\Delta s = v \Delta t = 2 d v/c_{\rm T}$. Using $v=0.1 \ {\rm m/s}$,
$c_{\rm T}=  56 \ {\rm m/s}$ and $d=8.6 \ {\rm nm}$ this gives $\Delta s = 0.031 \ {\rm nm}$.
%The negative values for
%$F_x/F_z$ occur when .. .
The numerical calculations show that the period (in time) of the oscillations in $F_x$ is
independent of the sliding speed $v$, and the contact pressure $p$, while the amplitude of the oscillations is proportional to
$v$.

For the commensurate interface the friction is much larger and
nearly velocity independent (see Fig. \ref{Fig.6a-commensurate-com-0.1-1.pdf}). This is the expected result
when at the sliding interface rapid slip events occur, involving velocities independent of the driving speed $v$.
Observations of movies (see Ref. \cite{Movies})
shows that the sliding motion involves domain-wall excitation (solitones), which propagate
with high speed (of order the Rayleigh sound velocity; see Fig. \ref{commensurate-sliding-detail.pdf}),
unrelated to the sliding speed $v$, while energy is radiated into the block giving rise to the observed high
friction force.

For both the commensurate and the incommensurate systems the contact width does not change with sliding speed in the studied
velocity range ($v<1 \ {\rm m/s}$).
For the incommensurate system this is expected because of the nearly vanishing (average) friction force, but for the commensurate
system the friction is large but still the contact width is independent of the sliding speed. In particular, there is no change
in the contact width between $v=0$ with $F_x=0$ and sliding at a finite velocity where the frictional shear stress is of order
$0.1 \ {\rm MPa}$ (as is typical for PDMS sliding on a glass surface\cite{France}).

In an earlier publication we have studied sliding friction for the flat-curved situation
where an elastic block with a smooth surface is sliding on a rigid surface with
the height profile $z=h_0 {\rm cos} (q_0 x)$. This case differs from the curved-flat configuration studied above since for
the flat-curved system there is an important contribution to the friction from phonon emission from the opening and closing crack tips.
In fact, for the system sizes we have studied, for the flat-curved case, even for the commensurate interface
the contact edge contribution to the friction is larger than the
contribution from the internal region of the contact. This is illustrated in Fig.
\ref{Fig.8-shear-stress-comm1x1-bendsub-0.1MPa-0.1ms.pdf} which shows the nominal frictional shear stress $F_x/(L_xL_y)$
as a function of the sliding distance for the flat-curved
case with incommensurate interface (red curve), and for the commensurate interface (blue curve).
In the calculations we have assumed the nominal contact pressure $p=0.1 \ {\rm MPa}$.
The figure also shows the result for a higher contact pressure, $p=0.3 \ {\rm MPa}$ (green line).
Note that the incommensurate interface gives the largest friction. For this system there is negligible contribution
to the friction from the internal region of the contact. That is,
the friction is entirely due to the phonon emission associated with the rapid atomic snap-out and
snap-in at the crack edges (see Ref. \cite{a3} and \cite{a6}).
It is remarkable that the commensurate system gives lower friction than the incommensurate system,
in spite of the fact that for this system
there is both a contribution from the internal region of the contact and from the crack edges. However, the contribution
from the crack edges is smaller than for the incommensurate system due to the higher density of substrate atoms
for the incommensurate system (the ratio is $206/128 \approx 1.61$).

Fig. \ref{Fig.9-Shear.stress-comp-comm-slid-0.1MPa-0.1ms-z.x.pdf}
shows (a) the normal stress $\sigma_{zz}$ and (b) the shear stress $\tau_{zx}$
acting on the block as a function of the spatial coordinate $x$. The red lines are for the curved-flat system,
and the green lines for the flat-curved system, in both cases with a commensurate interface (with $a_{\rm b}/a_{\rm s} = 128/128=1$).
The sliding speed $v=0.1 \ {\rm m/s}$, and the nominal contact pressure $p=0.1 \ {\rm MPa}$.
Note that in both cases at the edge of the contact region the normal stress $\sigma_{zz}$ is tensile and maximal, as expected
from the JKR theory, and from the theory of cracks, which predict that the stress has a $r^{-1/2}$ singularity
at $r=0$ (where $r$ is the distance from the crack tip). Because of the curved contact region in the flat-curved
system, at the edge of the contact region the stress $\tau_{zx}$ will exhibit a similar singular form as the normal stress.
However for the curved-flat system the contact region is flat and only the $\sigma_{zz}$ stress exhibit the
singular form.

\vskip 0.3cm
\centering{\bf DISCUSSION}
\justify
We have shown above that within linear elasticity theory the contact area
between an elastic cylinder and a rigid flat countersurface does not depend on the
applied tangential force, at least not for the systems studied above.
This is in contrast to some experimental results for PDMS spheres sliding on smooth
glass surfaces. This indicates that the origin of the area reduction in the size of the contact area
in Ref. \cite{Chaud} and \cite{PNAS,PRL,Cia} may be due to some effect not taken into account in the model study, such as
material viscoelasticity, elastic nonlinearity or contact time-dependent work of adhesion.

In an interesting study Lengiewicz et. al.\cite{a1} have found that
the observed contact area reduction for a PDMS rubber sphere in contact with a glass surface
can be explained by a theory based on non-linear elasticity without invoking adhesion!
They found quantitative agreement with the recent experimental results of Sahli
et. al. \cite{PNAS,PRL,a1} on sphere-plane elastomer contacts, without adjustable
parameters, using the neo-Hookean hyperelastic model.
The importance of elastic nonlinearity for the explanation of the contact area reduction
have been suggested by us in some earlier papers in particular the
Tribology Letters\cite{a2}. In another paper \cite{a3} we observed
that for a dry clean human finger there is no macroscopic adhesion to a flat glass
plate (due to the large surface roughness of the skin)
so the fact that for this case too the contact area is reduced upon application
of a tangential force must clearly be a non-linear elastic effect.
That adhesion itself may not result in a reduction in the contact area upon
application of a tangential force was also shown theoretically to be the case
in the paper by Menga, Carbone and Dini\cite{Menga}.
Finally, we note that in at least one study the rubber-glass contact area
was found to increase upon sliding indicating that other mechanisms may contribute to
the dependency of the contact area on the tangential force.\cite{Krick}

If the reduction in the contact area upon application of a tangential force
can be interpreted as an effective stiffening of the rubber elastic properties
by the tangential deformations, then we expect also that the penetration is decreasing
by increasing tangential force. This could explain the experimental observation
in Ref. \cite{a2} that a sharp tip indented in a rubber surface with a given normal force moves
upwards when a parallel force is applied in addition to the normal force.
Within (small deformation) linear elasticity theory this result is unexpected as there
is no coupling between the parallel and the perpendicular deformations when the Poisson ratio
is equal to 0.5 (incompressible solid), as is the case in the rubbery region
for rubber-like materials.

In this paper, and in Ref. \cite{a3} and  \cite{a6}, we have assumed simple crystalline solids.
Rubber-like materials are more complex materials with cross-linked long-chain molecules,
and may have nanometer-thick surface layers with liquid-like mobility of the
polymer segments, which can rearrange in the substrate surface potential and form
small regions which are pinned by the substrate. In this case during lateral
motion of the rubber block the chain stretches, detaches, relaxes, and reattaches to the surface to repeat the cycle.
Here ``detaches'' stands for rearrangement of molecule segments (in small domains) parallel
to the surface from pinned (commensurate-like) to depinned (incommensurate-like) domains.
This result in an ``area-dominated friction'' where the shear stress is uniform
within the contact area as observed experimentally\cite{France}. In this case the  friction force
arises from stick-slip type of motion of nanometer-sized regions everywhere within the
contact region. Theoretical model studies of this process was presented in Ref. \cite{Schall,smooth}.

\begin{figure}
\centering
\includegraphics[width=0.45\textwidth]{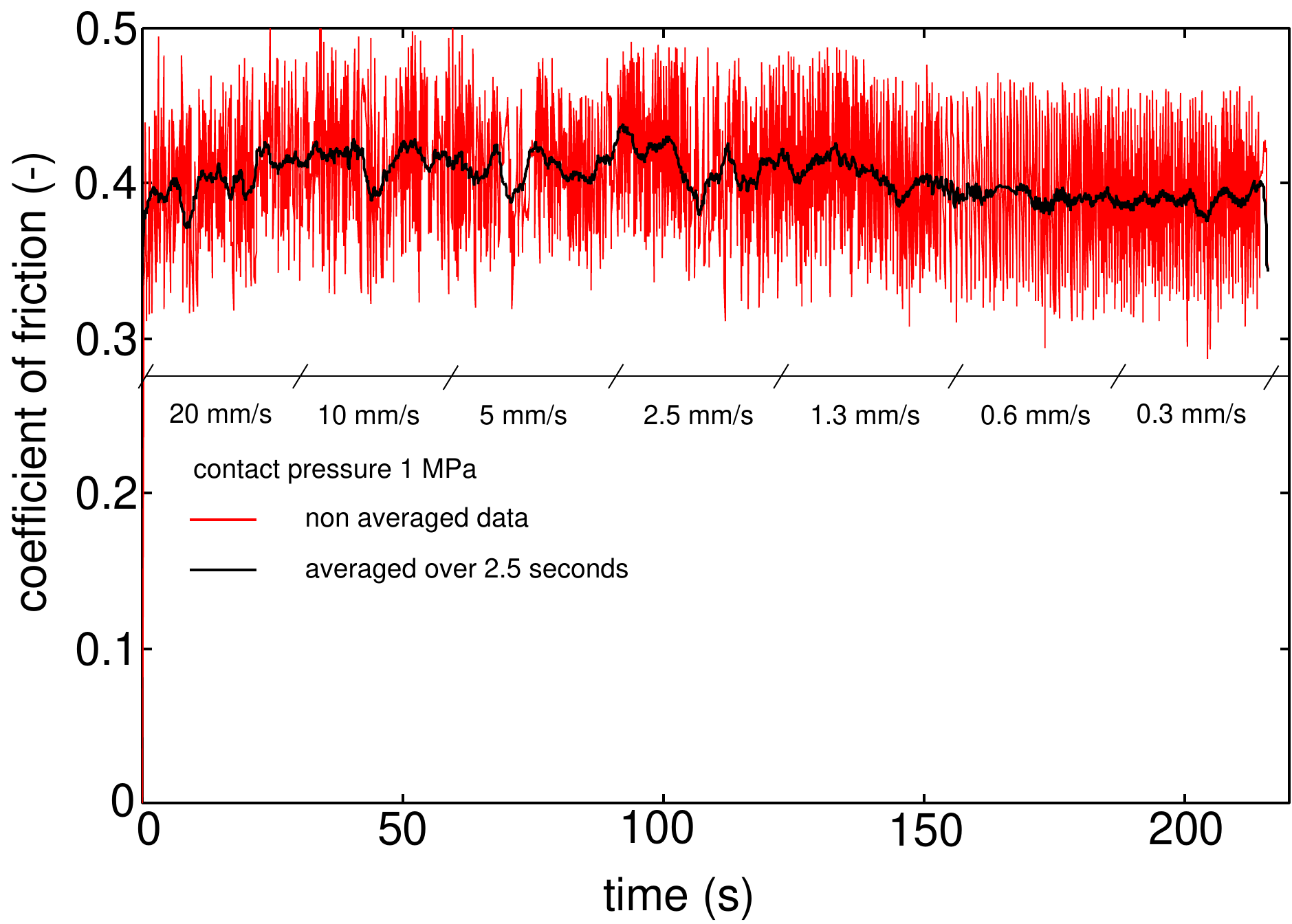}
\caption{\label{commensurate.pdf}
The sliding friction for
PolyOxyMethylene (POM) polymer block sliding in a POM substrate. For nominal flat interface
with the nominal contact pressure $1 \ {\rm MPa}$ and the temperature $T= 20^\circ {\rm C}$.
}
\end{figure}

That systems with (nearly) incommensurate interface exhibit smaller friction than systems with commensurate interface, as found above, is a well known, but in practice there are several ``complications''.
Thus, even if the interface is incommensurate mobile adsorbed contamination molecules will always exist in the normal
atmosphere, which will adjust their positions and pin the surfaces together, which may result in a
large and nearly velocity independent friction force\cite{He, Mus}. Nevertheless, even if there are no strictly incommensurate
systems, one expect smaller friction force the closer an interface becomes to a perfect incommensurate system.
As an example, if two polymers, say A and B,  with very different nature of the bead units, slide on top of each other,
a relative small friction coefficient may prevail, while for A sliding on A, or B on B, the friction may
be much higher due to the more commensurate-like contact\cite{Ion, Ion2}. In the latter cases the friction coefficient is also expected
to be nearly velocity independent, assuming the sliding speed is not so high that frictional heating becomes important, or so low
that thermal activation becomes important. As an example, in Fig. \ref{commensurate.pdf} we show the velocity dependency
for PolyOxyMethylene (POM) polymer block sliding on a POM substrate.

\vskip 0.3cm
\centering{\bf SUMMARY AND CONCLUSIONS}
\justify
We have presented molecular dynamics
simulations for an elastic cylinder sliding on a rigid flat countersurface (curved-flat).
For this system, within linear elasticity theory, the contact area does not depend on the applied
tangential force.

The sliding friction of commensurate and incommensurate interface contacts was
investigated. For the commensurate interface the friction is large and
nearly velocity independent due to rapid slip events involving domain-wall excitation (solitones),
which propagate at the interface with a speed of order the sound velocity, radiating energy to the block
giving rise to the observed high friction force.

The geometry of contact, whether curved-(rigid)flat contact or flat-(rigid)curved contact decide the
energy dissipation mechanism during sliding resulting in observed difference in frictional force.
For the curved-flat case the friction force is mainly due to processes occurring inside the contact area, while for the
flat-curved case, for the system we studied, the friction force is mainly due to phonon emissions at
the crack edges associated with the rapid atom
snap-out (at the opening crack) and snap-in (at the closing crack) events.

\vskip 0.3cm
\centering{\bf ACKNOWLEDGEMENTS}
\justify
 J. Wang would like to thank scholarship from China Scholarship Council (CSC) and funding by National Natural Science Foundation of China (NSFC): grant number U1604131.

\end{document}